\documentclass{PoS}

\title{Photon cluster excitation mechanism in an $\mathbf{e^{-}e^{+}}$-plasma
created from vacuum by a strong electromagnetic field}

\ShortTitle{The photon cluster excitation mechanism}

\author{\speaker{S.A.~Smolyansky}\\ 
        Saratov State University, Saratov, Russia\\
        E-mail: \email{smol@sgu.ru}}

\author{D.B.~Blaschke\\
        Institute for Theoretical Physics, University of Wroc{\l
}aw, Wroc{\l }aw, Poland\\
	Bogoliubov Laboratory for Theoretical Physics,
JINR Dubna, Dubna, Russia\\
        E-mail: \email{david.blaschke@gmail.com}}

\author{V.V.~Dmitriev\\
        Saratov State University, Saratov, Russia\\
        E-mail: \email{dmitrievv@gmail.com}}

\author{B.~K\"ampfer\\
        Helmholtz-Zentrum Dresden-Rossendorf, Dresden, Germany\\
        E-mail: \email{b.kaempfer@hzdr.de}}

\author{A.V.~Prozorkevich\\
        Saratov State University, Saratov, Russia\\
        E-mail: \email{avproz@bk.ru}}

\author{A.V.~Tarakanov\\
        Saratov State University, Saratov, Russia\\
        E-mail: \email{tarakanovav@sgu.ru}}

\abstract
{In the low density approximation of nonperturbative kinetic
theory  we investigate the vacuum creation of an
electron-positron plasma (EPP) under the action of an external
monochromatic "laser" electric field with the focus on multiphoton processes.
We analyse in some detail the role of the photon cluster mechanism of
EPP excitation and show that this mechanism plays the dominating role
in the case when a large photon number from the external field reservoir
takes part in the breakdown of the energy gap.
We obtain the distribution function in this approximation and discuss its
properties.}

\FullConference{XXI International Baldin Seminar on High Energy Physics
Problems,\\
		September 10-15, 2012\\
		JINR, Dubna, Russia}

\usepackage{graphicx,pstricks}
\usepackage{amsmath,amssymb}
\usepackage{amsfonts}

\begin{document}

\section{Introduction}

For the description of the vacuum creation of an electron-positron plasma (EPP)
under the action of a strong time dependent external electric field (the
dynamical Schwinger effect) the reliable tool is the kinetic equation (KE)\
\cite{1}.
It is an exact non-perturbative consequence of the basic equations
of motion of QED for a linearly polarized spatially homogeneous electric
field ("laser field").
In the present work we continue to analyse a quadrature solution of this KE\
obtained in the low density approximation \cite{2}.
We estimate here the role of photon cluster processes introduced \cite{3} in
comparison with the orthodox multiphoton mechanism and show that this
mechanism plays the dominant role in the case of a large photon number
absorbed from the external field reservoir.
It corresponds to the optical and X-ray range of the laser radiation.
The quasiparticle EPP excitation within a period of the laser pulse action
was investigated rather detailed (e.g., \cite{3a}).
Below we discuss the mechanisms of the residual EPP generation.

We restrict ourselves here to the consideration of the multiphoton domain
corresponding to large adiabaticity parameters
$\gamma =  (E_{c}/E_{0})\cdot(\nu/m) \gg 1$ and subcritical fields
$E_{0}\ll E_{c}=m^{2}/e$ .
The basic equations are represented in Sect.~2.
Corresponding to the multiphoton mechanism of the EPP\ excitation the
nonlinear harmonic analysis is discussed in Sect.~3.
Sect.~4 is devoted to an investigation of the momentum spectrum of the problem
and the threshold conditions of the energy gap breakdown.
The proof of the dominating role of the photon cluster process in the domain
of large total photon numbers is given in Sect.~5.
The distribution function in this approximation is obtained here also.
The basic results of the work are summarized in Sect.~6.

\section{Low density limit
}

For the description of the EPP vacuum creation we use the exact
nonperturbative KE\ obtained in the work \cite{1} for an arbitrary time
dependent and spatially homogeneous electric field with linear polarization.
In Hamiltonian gauge, $A^{\mu }(t)=\left( 0,0,0,A(t)\right) $,
the electric field strength is $E(t)=-\dot{A}(t)$.
The one-body electron (positron) phase space distribution is
\begin{equation}
\dot{f}(\mathbf{p},t)=\frac{1}{2}\lambda (\mathbf{p},t)\int%
\limits_{t_{0}}^{t}dt^{\prime }\lambda (\mathbf{p},t^{\prime })\left[ 1-2f(%
\mathbf{p},t^{\prime })\right] \cos \theta (t,t^{\prime }),  \label{1}
\end{equation}%
where
\begin{equation}
\lambda (\mathbf{p},t)=eE(t)\varepsilon _{\bot }/\omega ^{2}(\mathbf{p},t)
\label{2}
\end{equation}%
denotes the amplitude of the EPP excitation,
$\omega (\mathbf{p},t)=
\sqrt{\varepsilon _{\bot }^{2}(\mathbf{p})+(p_{\parallel }-eA(t))^{2}}$
stands for the quasienergy with the transverse energy
$\varepsilon _{\bot }=(m^{2}+p_{\perp}^{2})^{1/2}$,
and the high frequency phase is
\begin{equation}
\theta (t,t^{\prime })=2\int_{t^{\prime }}^{t}d\tau \omega
(\mathbf{p},\tau ) .
\label{3}
\end{equation}
The distribution function in the quasiparticle representation is defined
for the in-vacuum state,
$f(\mathbf{p},t)= \langle in|a^{+}(\mathbf{p},t) a(\mathbf{p},t)|in \rangle$.
For the generalization of the KE (\ref{1}) to arbitrary
electric field polarization see \cite{4}-\cite{6}.

In the low-density limit, $f\ll 1$, KE\ (\ref{1}) obeys the formal
solution \cite{2}
\begin{equation}
f_{\rm low}(\mathbf{p},t)=\frac{1}{4}\left\vert \int\limits_{-\infty
}^{t}dt^{\prime }\lambda (\mathbf{p},t^{\prime })e^{i\theta (t,t^{\prime
})}\right\vert ^{2}~,
\label{4}
\end{equation}
were the external field is supposed to be switched on in the infinite past,
$t_{0}\rightarrow -\infty $.
The low-density approximation implies that the external field is weak, i.e.,
$E\ll E_{c}$.
Below we will restrict ourselves to the
analysis of Eq. (\ref{4}) in the limit $t\rightarrow \infty $, which defines
the momentum distribution of the residual EPP [3],
\begin{equation}
f_{out}(\mathbf{p})=\lim_{t\rightarrow \infty }f_{low}(\mathbf{p},t)=
\frac{1}{4}\left\vert \int\limits_{-\infty }^{\infty }dt
\lambda (\mathbf{p},t)e^{i\theta (t)}\right\vert ^{2}.
\label{5}
\end{equation}
(To arrive at (\ref{5}) the representation of the phase (\ref{3}) via
antiderivatives has been used,
$\theta (t,t^{\prime })=\theta (t)-\theta (t^{\prime})$.)

Hereafter we will consider the domain of the multiphoton mechanism
of EPP vacuum creation, where the adiabaticity parameter defined above is
large [8], $\gamma \gg 1 $.
A perturbation theory in the small parameter $1/\gamma \ll 1$ can
be constructed here.
In the case of a monochromatic laser field
\begin{equation}
A(t)=(E_{0}/\nu )\cos \nu t,\qquad E(t)=E_{0}\sin \nu t,
\label{6}
\end{equation}
this statement is based on the representation of the quasimomentum
\begin{equation}
P=p_{\parallel }-\frac{m}{\gamma }\cos \nu t.
\label{7}
\end{equation}

As a first step let us consider (\ref{5}) in an expansion.
In leading order, one obtains
\begin{equation*}
\omega (\mathbf{p},t)\quad \rightarrow \quad \omega _{0}(\mathbf{p})
=\sqrt{\mathbf{p}^{2}+m^{2}}~,
\end{equation*}
\begin{equation}
\lambda (\mathbf{p},t)\quad \rightarrow \quad \lambda _{0}(\mathbf{p})E(t),
\quad
\lambda _{0}(\mathbf{p})=e\varepsilon _{\perp }/\omega _{0}^{2}(\mathbf{p}),
\label{8}
\end{equation}
and $\theta (t)= 2 \omega _{0}\, t$.
This means that contributions of the high frequency harmonics of the external
field are neglected in Eq.~(\ref{5}), yielding the result
\begin{equation}
f_{out}(\mathbf{p})=
\frac{1}{4}\lambda _{0}^{2}\left\vert E(\omega =2\omega_{0})\right\vert ^{2},
\label{9}
\end{equation}
where $E(\omega )$ is the Fourier transform of the field strength $E(t)$.
In this approximation, the vacuum EPP production takes place
when for the frequency of the one-photon $e^{-}e^{+}$ pair creation
process $\omega _{1\gamma }$ holds $\omega _{1\gamma }=2\omega_{0}$.
This mechanism is exclusive here.
Its intensity is regularized by the presence of the frequency
$\omega _{1\gamma }$ in the spectrum of an external field.

Let us consider the special case of a monochromatic field (\ref{6}).
The distribution function $f(\mathbf{p})$ becomes
\begin{equation}
f_{out}(\mathbf{p})=\frac{1}{4}\lambda _{0}^{2}E_{0}^{2}\delta \left[
2\omega _{0}(\mathbf{p})-\nu \right] \delta (0)~.
\label{9a}
\end{equation}
Substituting $\delta (0)\rightarrow \delta (2\pi /T_{p})\rightarrow
T_{p}/2\pi $
(where $T_{p} \gg 2\pi/\nu$ is the period of the field pulse action),
we obtain the spectral density for the production of the EPP
\begin{equation}
\frac{df_{out}(\mathbf{p})}{dt}=\frac{1}{8\pi }\lambda
_{0}^{2}E_{0}^{2}\delta \left[ 2\omega _{0}(\mathbf{p})-\nu \right] .
\label{9b}
\end{equation}
The $\delta $ distribution defines here the admissible momentum set for $\nu
\geqslant 2m$. Thus, employment the field (\ref{6}) nonconvergent for
$t\rightarrow \pm \infty $ leads to the singular distributions (\ref{9a})
and (\ref{9b}), for which the physical meaning is restored by calculating
physical quantities such as densities.

In order to facilitate breakdown of the energy gap by EPP creation, one can
take into account the multiphoton mechanism of EPP production. These
processes are described by the high frequency multiplier in Eq.~(\ref{5}).
The relevant methods will be developed in the next Section.

\section{Multiphoton processes}

Let us use the non-perturbative method of photon counting assuming
that the electric field is periodical $A(t)=A(t+2\pi /\nu )$, where
$\nu =2\pi /T$ is the angular frequency and $A(t)=A(-t)$, in order
to provide the property of
being an even function of the quasiparticle energy,
$\omega (\mathbf{p},t)=\omega (\mathbf{p},-t)$.
The corresponding Fourier transform
\begin{equation}
\omega (\mathbf{p},t)=\sum_{n=0}\Omega _{n}\cos n\nu t  \label{10}
\end{equation}
leads to the decomposition of the phase in Eq.~(\ref{5})
\begin{equation}
\theta (t)= 2 \Omega _{0}t+\sum_{n=1}a_{n}\sin n\nu t~,
\label{11}
\end{equation}
where $a_{n}=2\Omega _{n}/\nu n$ .
In case of a harmonic external field, $\Omega _{0}$ is the renormalized
frequency \cite{7}.
Let us employ now the approximation
(\ref{8}) and use the non-perturbative decomposition
\begin{equation}
\exp (ia\sin \phi )=\sum_{k=-\infty }^{\infty }J_{k}(a)e^{ik\phi },
\label{12}
\end{equation}%
where $J_{k}(a)$ are the Bessel functions of order $k$.
Let us now consider the integral
in Eq.~(\ref{5}) and
perform the substitutions (\ref{10})-(\ref{12}), leading to
\begin{eqnarray}
J &=&\int\limits_{-\infty }^{\infty }dtE(t)e^{i\theta (t)}  \notag \\
&=&\int\limits_{-\infty }^{\infty }dtE(t)e^{2i\Omega _{0}t}\left\{
J_{0}(2\Omega _{0})+\prod\limits_{n=1}\sum_{k=1}J_{-k}(a_{n})e^{-i(kn)\nu
t}+(k\leftrightarrow -k)\right\} ~.
\label{13}
\end{eqnarray}

The first term with $J_{0}(2\Omega _{0})$ describes the direct vacuum
excitation at the frequency $\omega _{1\gamma }=2\Omega _{0}$ and
corresponds to Eq.~(\ref{9a}).
If we are interested in excitations at lower frequencies, this contribution
can be omitted.
One can omit in Eq.~(\ref{13}) also the series with the substitution
$k\rightarrow -k$ since the corresponding frequencies cannot lead to a
compensation of the high frequency phase $2\Omega _{0}t$.
At last, let us replace the infinite series and products by finite ones
to arrive at the expression
\begin{equation}
J(\mathbf{p},N_{ph},N_{c})=\int\limits_{-\infty }^{\infty }dtE(t)e^{2i\Omega
_{0}t}\prod\limits_{n=1}^{N_{ph}}\sum_{k=1}^{N_{c}}J_{-k}(a_{n})e^{-i(kn)\nu
t}.  \label{14}
\end{equation}

Here the multiphoton processes of the energy absorption from the photon
reservoir of the external field are characterized by the pair of indices $n$
and $k$.
The index $n$ corresponds to the ordinary multiphoton process (see its source
in Eqs.~(\ref{10}), (\ref{11})), whereas the index $k$ marks the group
(cluster) multiphoton process when each of $k$ clusters contains $n$ identical
photons with the energy $\nu $.
The ordinary multiphoton process corresponds to the simplest "cluster" of the
order $k=1$.
The appearance of photon clusters in the multiphoton processes in
Eq.~(\ref{14}) is a consequence of the nonlinear field dependence of the
quasiparticle energy $\omega (\mathbf{p},t)$.
The probability for the generation of a $n$-photon cluster is defined by the
amplitude $a_{n}$ in the decomposition (\ref{11}) (the argument of the Bessel
function $J_{k}(a_{n})$ in Eq.~(\ref{14})) while the probability of a
simultaneous $k$-cluster absorption corresponds to the Bessel function of the
order $k$.

Now one can identify the numbers $N_{ph}$ and $N_{c}$ in Eq.~(\ref{14}):
$N_{ph}$ is the maximum number of photons in a cluster, and $N_{c}$ is
the maximum number of clusters.
The limit $N_{ph}\rightarrow \infty $, $N_{c}\rightarrow \infty $ corresponds
to the exact formula (\ref{13}).

In order to perform the integration over $t$, it is necessary to carry out
 the summation in Eq.~(\ref{14}).
For this aim let us remark that products of the different series in
Eq.~(\ref{14}) among themselves generate the exponent product.
Thus, for the simplest harmonic field (\ref{6}) we obtain
\begin{eqnarray}
J(\mathbf{p},N_{ph},N_{c}) &=&-i\pi E_{0}\sum_{k=1}^{N_{c}}\left\{ \delta
(2\Omega _{0}-\nu k\mathcal{N}_{ph}+\nu )
-\delta (2\Omega _{0}-\nu k\mathcal{N}_{ph}-\nu )\right\}
\prod\limits_{n=0}^{N_{ph}}J_{-k}(a_{n})~,
\label{15}
\end{eqnarray}
where the total photon number in a cluster of first order ($k=1$) is
\begin{equation}
\mathcal{N}_{ph}=\sum_{n=1}^{N_{ph}}n=\frac{1}{2}N_{ph}(N_{ph}+1)~.
\label{16}
\end{equation}
Let us substitute now the relation (\ref{15}) into Eq.~(\ref{5}).
This leads to a sum of products of two $\delta $-distributions.
They refer to (i) the energy conservation law in the multiphoton process
\begin{equation}
2\Omega _{0}(\mathbf{p})-\nu N_{tot}^{\pm }=0,
\label{17}
\end{equation}
where $N_{tot}^{\pm }=k\mathcal{N}_{ph}\pm 1$ is the total number of
the photons taking part in the process (contribution $\pm 1$ is
stipulated by the multiplier $E(t)$ in Eq.~(\ref{14})), and
(ii) to the argument of the cluster processes in the product $JJ^{\ast }$
leading to either $k_{1}=k_{2}=k$ or $(k_{1}-k_{2})\mathcal{N}_{ph}=\pm 2$.
The last condition can not be fulfilled for any integer numbers $k_{1}$ and
$k_{2}$.
Thus, we obtain for the distribution function (\ref{5})
\begin{eqnarray}
f_{out}(\mathbf{p},N_{ph},N_{c}) &=&
\frac{\pi ^{2}\lambda _{0}^{2}E_{0}^{2}}{4\nu \mathcal{N}_{ph}}
\sum_{k=0}^{N_{c}}\left\{ \delta \lbrack 2\Omega_{0}-\nu N_{tot}^{+}]+
+\delta \lbrack 2\Omega _{0}-\nu N_{tot}^{-}]\right\}
\prod\limits_{n=0}^{N_{ph}}J^2_{-k}(a_{n})~.
\label{18}
\end{eqnarray}
Here, we used the relation
\begin{equation}
\sum_{k^{\prime }}\delta \lbrack \nu \mathcal{N}_{ph}(k-k^{\prime})]
F_{k,k^{\prime }}=\frac{1}{\nu \mathcal{N}_{ph}}F_{k,k}
\label{19}
\end{equation}
for an arbitrary function $F_{k,k^{\prime }}$.
Such an approach allows to avoid the hypothesis of a phase randomization [3].

\section{Spectrum}

Our next goal is the investigation of the equation (\ref{17}), which defines
the threshold photon numbers and the spectrum of the created EPP.
For a given frequency $\nu $, the conditions (\ref{17}) fix the minimum
combinations
\begin{equation}
N_{c}^{tr}\mathcal{N}_{ph}^{tr}=2\Omega _{0}(\mathbf{p}=0)/\nu \pm 1
\label{20}
\end{equation}
of the threshold photon ($N_{ph}^{tr}$) and cluster ($N_{c}^{tr}$) numbers,
whereby the breakdown of the energy gap is possible for the 
produced EPP.
Absorption of a greater photon number with
$k\mathcal{N}_{ph}>N_{c}^{tr}N_{ph}^{tr}$
generates an EPP with $p\neq 0$.

The first step in the analysis of the threshold conditions is the determination
of the momentum dependence of the renormalized quasienergy
$\Omega _{0}(\mathbf{p})$ in explicit form.
According to Eq.~(\ref{10})
\begin{equation}
\Omega _{0}(\mathbf{p})=\frac{\nu }{2 \pi }\int_{-\pi /\nu }^{\pi /\nu
}\omega (\mathbf{p},t)dt=\frac{1}{\pi }\int_{-1}^{1}du\left\{ \frac{%
\varepsilon _{\perp }^{2}+(p_{\parallel }- u\, m / \gamma)^{2}}{1-u^{2}}\right\}
^{1/2}  \label{21}
\end{equation}%
for the field (\ref{6}). From here it is possible to receive various approximations. The upper bound is
\begin{equation}
\Omega _{0}(\mathbf{p})=
\sqrt{\omega _{\ast }^{2}(\mathbf{p})- 2 p_{\parallel} m /\gamma}~,
\label{22}
\end{equation}
where $\omega_{\ast }(\mathbf{p})$ is the renormalized quasienergy with the
mass $m_{\ast }= m \sqrt{1+1/\gamma^2}$ being renormalized in the field.
From Eq.~(\ref{22}) one sees that the anisotropy effect is proportional to
$1/\gamma $.
In the isotropic approximation we obtain
$\Omega _{0}(\mathbf{p})=\omega _{\ast }(\mathbf{p})$.

From Eqs.~(\ref{17}) and \ (\ref{22}) we find the momentum spectrum of the
created EPP
\begin{equation}
p_{1,2}/m_{\ast }=
\frac{1}{\gamma }\cos{\phi}
\pm \left\{ \left[ \frac{\nu}{2m_{\ast }}N^{\pm}_{tot}\right] ^{2}-1
+\frac{1}{\gamma ^{2}}\cos^{2}{\phi}\right\} ^{1/2}~,
\label{23}
\end{equation}
where $\phi $ is the angle between the vector $\mathbf{p}$ and the axis
$p^{3}=p_{\parallel }$.

Let us consider below the isotropic approximation.
The momentum spectrum (\ref{23}) transforms then according to
\begin{equation}
p=m_{\ast }\left\{ \left[ \frac{\nu }{2m_{\ast }}N_{tot}^{\pm }\right]
^{2}-1\right\} ^{1/2}~.
\label{24}
\end{equation}
For $p=0$ the threshold condition follows from here (the symbol $[x]$ means
the integral part of $x$)
\begin{equation}
\frac{1}{2}N_{c}^{tr}N_{ph}^{tr}(N_{c}^{tr}+1)\pm 1= [2m_{\ast }/\nu] ~,
\label{25}
\end{equation}
which defines the set of the threshold numbers $N_{c}^{tr}$ and
$N_{ph}^{tr}$ for the given total photon number that is necessary for a
breakdown of the energy gap.
For $N_{tot}^{+}=1$, the one-photon mechanism of EPP excitation is realized
again.
Combinations of the numbers $N^{tr}_{c}$ and $N^{tr}_{ph}$
(compatible with the threshold condition (\ref{25}))
are presented in Table 1 for the simplest few-photon processes with the
minimum number $N^{\pm}_{tot}$.
From this table one can see that  number  of  the  photon  degrees  of
freedom is larger than for conventional multiphoton processes, at least in the
domain of the few-photon mechanism, i.e. for $\nu \sim m$.

\begin{figure}
\begin{minipage}{0.5\textwidth}
\includegraphics[width=0.95\textwidth]{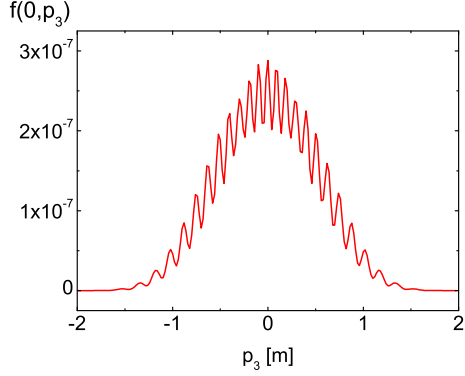}
\caption{A typical distribution function of the residual EPP in periodic field with Gaussian shape ($E=0.2 E_c, \lambda = 0.01 nm$) \label{fig1}}
\end{minipage}\hfill
\begin{minipage}{0.45\textwidth}
\begin{center}
{\bf Table 1:} Distribution of the total photon numbers $N_{\rm tot}$
to clusters\\[5pt]
\begin{tabular}{|c|c|c|}
\hline
$2m_{\ast }/\nu $ & {$N_{c}^{tr}$} &
{$N_{ph}^{tr}$} \\ \hline
$1=N_{tot}^{+}$ & 1 & 0 \\ \hline
$1=N_{tot}^{-}$ & 2 & 1 \\ \hline
$2=N_{tot}^{+}$ & 1 & 1 \\ \hline
$2=N_{tot}^{-}$ & 1 & 2 \\ \hline
$2=N_{tot}^{-}$ & 3 & 1 \\ \hline
$3=N_{tot}^{+}$ & 2 & 1 \\ \hline
$3=N_{tot}^{-}$ & 4 & 1 \\ \hline
$4=N_{tot}^{+}$ & 1 & 2 \\ \hline
$4=N_{tot}^{+}$ & 3 & 1 \\ \hline
$4=N_{tot}^{+}$ & 5 & 1 \\ \hline
\end{tabular}
\end{center}
\end{minipage}
\end{figure}

Let us introduce now the superthreshold photon ($\delta n$) and cluster
($\delta k$) numbers.
The corresponding total numbers will be equal $N_{ph}=N_{ph}^{tr}+\delta n$
and $N_{c}=N_{c}^{tr}+\delta k$.
The numbers $\delta n$ and $\delta k$ define the discrete momentum spectrum
\begin{equation}
p_{\delta n,\delta k}=m_{\ast }\sqrt{\frac{\nu }{m_{\ast }}}\delta
^{1/2}(N_{c},N_{ph})\left[ \frac{\nu }{4m_{\ast }}\delta (N_{c},N_{ph})
+1\right] ^{1/2},
\label{26}
\end{equation}
where
\begin{equation}
\delta (N_{c},N_{ph})=N_{c}^{tr}\delta \mathcal{N}_{ph}
+\mathcal{N}_{ph}^{tr}\delta k+\delta \mathcal{N}_{ph}\delta k~.
\label{27}
\end{equation}
Thus, we obtain a discrete spherical symmetrical (in the isotropical
approximation) momentum spectrum.
For $\nu \ll m$ and $\delta n\ll N_{ph}^{tr}$ and $\delta k\ll N_{c}^{tr}$,
one obtains
\begin{equation}
p_{\delta n,\delta k}= \{m_{\ast}\, \nu\, \delta (N_{c},N_{ph})\}^{1/2}~,
\label{28}
\end{equation}
i.e., the spectrum is quasicontinuous in this case.

Using the solution of Eq.~(\ref{17}), we can rewrite the $\delta $
distributions in Eq.~(\ref{18}) in the form
\begin{equation}
\delta \left[ 2\Omega _{0}-\nu N_{tot}^{\pm }\right]
=\frac{\omega _{\ast }}{2p_{\delta n,\delta k}}\delta(p-p_{\delta n,\delta k})
~.
\label{29}
\end{equation}

The next subsidiary problem is an analytical estimate of the Fourier
coefficients of the series (\ref{10})
\begin{equation}
\Omega _{n}=\frac{\nu }{\pi }\int_{-\pi /\nu }^{\pi /\nu }dt\
\omega (\mathbf{p},t)\cos n \nu t,\qquad n\geqslant 1
\label{30}
\end{equation}
and the corresponding arguments $a_{n}$ of the Bessel functions.
It can be shown that each Fourier coefficient (\ref{30}) can
be represented by a series expansion with respect to the adiabaticity
parameter $1/\gamma \ll 1$.
The minimum power of such a kind of decompositions is fixed by the harmonic
number, $\Omega _{n}\sim 1/\gamma ^{n}$.
It is important that the even harmonics only correspond to the isotropical
distribution in the momentum space.
As a result we obtain for the even harmonics
\begin{equation}
a_{n}=\frac{n-1}{n\cdot n!2^{2n-2}}\left( \frac{m}{\gamma \omega _{\ast }}
\right) ^{n}\frac{\omega _{\ast }}{\nu },\qquad n\geqslant
N_{ph}^{tr}=2s,~~s=1,2,...
\label{31}
\end{equation}
Using the definition $N_{tot}^{\pm }$, Eq.~(\ref{16}) and Table 1 (for small
numbers $n$ and $k$), we obtain for $\Omega_{0}=\omega$
\begin{equation}
\frac{\omega _{\ast }}{\nu }=\frac{1}{2}\left[ \frac{1}{2}kn(n+1)\pm 1\right]
\label{32}
\end{equation}
{and hence $a_{n}^{\pm } \gg 1$ for large photon numbers and
$a_{n}^{\pm }\ll 1$ for small ones.}
This allows to use the asymptotic representation of the Bessel functions for
small or large values of the arguments.

\section{Role of the photon cluster processes}

The following working formula follows from Eqs.~(\ref{18}) and (\ref{29})
\begin{equation}
f_{out}(\mathbf{p};N_{ph},N_{c})
=\frac{\pi e^{2}E_{0}^{2}\varepsilon _{\perp}^{2}}{2 \nu \mathcal{N}_{ph}
\omega _{\ast }^{3}~p}\sum_{k=0}^{N_{c}-N_{c}^{tr}}
\delta (p-p_{\delta n,\delta k})
\prod\limits_{n=0}^{N_{ph}-N_{ph}^{tr}}J_{N_{c}^{tr}+k}^{2}(a_{N_{ph}^{tr}+n}).
\label{35}
\end{equation}
Here, the even harmonics are considered only.

In order to use the formula (\ref{35}), it is necessary to find the
threshold numbers $N_{ph}^{tr}$ and $N_{c}^{tr}$.
Equation (\ref{25}) serves for the definition of them.
For a fixation of these numbers the following prescription is offered:\
for a given frequency $\nu $ the different orders $N_{c}^{tr}$ of clusters
are tested and for each number $N_{c}^{tr}$ the corresponding photon number
$N_{ph}^{tr}$ is found.
These threshold numbers correspond to the creation of an EPP where all pairs
are at rest.
The superthreshold numbers with $\delta n\geqslant 0$ and $\delta k\geqslant 0$
describe the moving EPP.

Now we compare the efficiency of the usual multiphoton and cluster mechanisms.
The simplest method is a comparison of two limiting values of the distribution
function for $p=0$ corresponding to either the maximum photon number in one
cluster (the first limiting case, the ordinary multiphoton process) or the
maximal cluster number with the minimal photon number ($n=2$) in each
cluster (the second limiting case) with the same total threshold photon
number in both cases. Let us assume that the total photon number is large
\begin{equation}
N_{tot}^{\pm }=N_{tot}=k\mathcal{N}_{ph}=\frac{k}{2}N_{ph}(N_{ph}+1)\gg 1.
\label{36}
\end{equation}
Then we have (i) in the multiphoton case
\begin{equation}
k=1,\quad N_{tot}=N_{ph}^{2}/2;\quad N_{ph}^{tr}=2\sqrt{[m_{\ast }/\nu ]}~,
\label{37}
\end{equation}
and (ii) in the multicluster case
\begin{equation}
N_{ph}^{tr}=2~\text{(the~second~harmonic~only), }N_{c}^{tr}
=N_{tot}/3=[2m_{\ast }/3\nu ]~.
\label{38}
\end{equation}
Basing on Eq.~(\ref{35}), we rewrite the distribution function for
the threshold values of the photon numbers as
\begin{equation}
f_{out}(\mathbf{p};N_{ph}^{tr},N_{c}^{tr})=
\frac{\pi e^{2}E_{0}^{2}\varepsilon_{\perp }^{2}}{2 \nu \mathcal{N}_{ph}
\omega _{\ast}^{3}~p}J_{N_{c}^{tr}}^{2}(a_{N_{ph}^{tr}})\delta (p)~.
\label{39}
\end{equation}

In order to avoid the infrared divergence problem, we introduce the
clustering coefficient defined as the ratio of two limited distribution
functions: the multicluster ($f_{c}$) and multiphoton ($f_{ph}$)
distribution functions with the corresponding threshold numbers (\ref{37})
and (\ref{38})
\begin{equation}
\xi =f_{c}/f_{ph}~.
\label{40}
\end{equation}
Substitution of Eqs.~(\ref{31}) and (\ref{39}) leads to the strong inequality
\begin{equation}
\xi \gg 1~.
\label{41}
\end{equation}
Thus, the photon cluster mechanism is dominating in the region of large
total photon numbers.

This conclusion allows to generalize Eq.~(\ref{39}) for the case of an EPP
with pairs of nonzero momentum, that corresponds to the superthreshold photon
cluster numbers $N_{c}\gg N_{c}^{tr}$,
\begin{equation}
f_{out}(\mathbf{p};N_{ph}^{tr}=2,N_{c}^{tr})=\frac{\pi
e^{2}E_{0}^{2}\varepsilon _{\perp }^{2}}{6 \nu  \omega _{\ast
}^{3}~p}\sum_{k=0}^{N_{c}-N_{c}^{tr}}J_{N_{c}^{tr}+k}^{2}(a_{2})\delta
(p-p_{\delta n,\delta k}),  \label{42}
\end{equation}
where according to Eq.~(\ref{31})
\begin{equation}
a_{2}=\frac{m_{\ast }^{2}}{16\gamma ^{2}\omega _{\ast }\nu }.  \label{43}
\end{equation}
The structure of Eq.~(\ref{42}) ensures a slow power (nonexponential)
decrease of the distribution function in the momentum space.
This distribution is very extensive (of the order of $m$) and rapidly
oscillating relative to some average value.

The transition to the superthreshold cluster numbers $N_{c}\gg N_{c}^{tr}$ in
Eq.~(\ref{42}) is reflected here in two effects: the appearance of the higher
cluster harmonics (index of the Bessel functions) and the account of EPP
motion ($p\neq 0$).
The first feature describes, apparently, the foliated structure of the
distribution function \cite{3} whereas the second one defines the momentum
dependence on it.
Below we will consider the last effect only.
Then, in this approximation, we obtain from Eq.~(\ref{42}) in the limit
$N_c \to \infty$
\begin{equation}
f_{out}(\mathbf{p})=
\frac{\pi e^{2}E_{0}^{2}\varepsilon _{\perp }^{2}}{6 \nu \omega _{\ast}^{3}~p}
\sum_{k=0}^{\infty} J_{N_{c}}^{2}(a_{2})\delta (p-p')~,
\label{46}
\end{equation}
where according to Eq.~(\ref{29}) $p'=2k$ that corresponds to two photons in a
cluster.
Substitution of the summation by an integration leads to the result
\begin{equation}
f_{out}(\mathbf{p}) =
\frac{\pi e^{2}E_{0}^{2}\varepsilon_{\perp }^{2}(p)}
{3 \nu \omega_{\ast}^{3}(p) ~p^2} J_{N_{c}^{tr}}^{2}[a_{2}(p/2)]~.
\label{47}
\end{equation}
According to Eqs.~(\ref{32}) and (\ref{43}), the relation $a_2 \gg 1$ holds.
Using the corresponding asymptotic representation of the Bessel functions
\begin{equation}\label{48}
    J_n(z) = \sqrt{2/\pi z}\, \cos{(z-\pi n/2  - \pi /4)}~, \quad z \gg 1~,
\end{equation}
one can obtain from Eq.~(\ref{47}) the distribution function in the considered
approximation as
\begin{equation}\label{49}
  f_{out}(\mathbf{p})= f^{sm}_{out}(\mathbf{p}) + f^{os}_{out}(\mathbf{p})~,
\end{equation}
where
\begin{equation}\label{50}
    f^{sm}_{out}(\mathbf{p}) =
\frac{16 e^{2}E_{0}^{2}\varepsilon _{\perp }^{2}  \gamma^2 }
{ 3 \omega _{\ast }^{3}   p  \, m^2_{\ast}}
\end{equation}
is the smooth part and
\begin{equation}\label{51}
    f^{os}_{out}(\mathbf{p}) = f^{sm}_{out}(\mathbf{p}) \cdot
\cos{\left[\frac{m^2_{\ast}}{8 \gamma^2 \omega _{\ast } \nu } -
\pi ( n + 1/2) \right]}
\end{equation}
is the oscillating part in the momentum space.

Thus, the simplest approximation (\ref{49})-(\ref{51}) reproduces reasonably
well some important features of the results of numerical calculations of the
distribution of the residual EPP on the basis of Eq.~(\ref{5}), see Fig. 1.

\section{Summary}

In the present work, we have developed further the ideas of \cite{3,8}
which initiated the consideration of the dynamical Schwinger process
as a multiphoton process.
Here, the new class of the multiphoton processes to wit the photon cluster
one was introduced.
Such a mechanism describes an other analytical tool in comparison to the
orthodox multiphoton processes and is stipulated by the nonlinear field
dependence of the quasienergy.
The photon cluster process is a new class of cooperative effects in which
the photon set behaves as one photon with the total energy of the set.

We analyzed the distribution function of the residual EPP in the
multiphoton region in the low density approximation.
The basic result of our work is proving the statement that the photon
cluster process is dominating if the total photon number is large.
Our preliminary analysis has shown also that the orthodox multiphoton
process plays the main role in contrast to the case of small total
photon numbers.

We had obtained the final formula for the distribution function of the
residual EPP in the region of action of the cluster mechanism EPP excitation
and noted some its features:

- the distribution function has cylindrical symmetry and is flattened in
the direction of the acting electric force;

- the distribution function is very extensive (of order $m$) and rapidly
oscillating in the momentum space.

Subsequent development of this research direction addresses has not only the
current problems (calculation of the number density EPP, problem of the
infrared divergence, estimation of the distribution function in the
dominant domain of the orthodox multiphoton mechanism etc), but also
more deep questions as well (e.g., understanding the character of the
rapid oscillations in momentum space (Fig.~1), correlation
between the quasiparticle and residual EPP).

\textbf{Acknowledgements }

The authors are indebted to S. Schmidt for discussions and collaboration
on different aspects of this work and acknowledge {\L}.~Juchnowski and A.~Otto
for correspondence on solutions of the kinetic equations.
S.A.S. acknowledges support by Deutsche Forschungsgemeinschaft \ (DFG) under
Project number TO 169/16-1;
The work of V.V.D. has been supported in part by the Federal
Special-Purpose Program \textquotedblleft Cadres\textquotedblright\
of the Russian Ministry of Science and Education under project No. 8174.
D.B. acknowledges support by the Polish Narodowe Centrum Nauki within the
``Maestro'' programme and by the Russian Fund for Basic Research under grant
number 11-02-01538-a.


\end{document}